\documentclass[12pt]{iopart}

\usepackage{graphicx}% Include figure files
\usepackage{dcolumn}% Align table columns on decimal point
\usepackage{bm}% bold math
\usepackage{multirow}

\newcommand{\dd}{{\textrm d}}
\newcommand{\GeV}{{\textrm{GeV}}}
\newcommand{\MeV}{{\textrm{MeV}}}

\begin{document}

\title{The Non-Perturbative Analytical Equation of State for the Gluon Matter: I}

\author{V. Gogokhia and M. Vas\'{u}th}

\vspace{3mm}

\address{HAS, CRIP, RMKI, Depart. Theor. Phys., Budapest 114,
P.O.B. 49, H-1525, Hungary}

\eads{\mailto{gogohia@rmki.kfki.hu}, \mailto{vasuth@rmki.kfki.hu}}

%\date{\today}

\begin{abstract}
The effective potential approach for composite operators is
generalized to non-zero temperatures in order to derive the equation
of state for pure $SU(3)$ Yang-Mills fields. In the absence of
external sources, this is nothing but
the vacuum energy density. The key element of this derivation is
the introduction of a temperature dependence into the expression
for the bag constant. The non-perturbative analytical
equation of state for gluon matter does not depend on the
coupling constant, but instead introduces a dependence on the mass gap. This is responsible for the
large-scale structure of the QCD ground state.
The important thermodynamic quantities, such as the
pressure, energy and entropy densities, etc., have been calculated.
We show explicitly that the pressure may vary continuously
around $T_c = 266.5 \ \MeV$, whereas all other thermodynamic
quantities undergo drastic changes at this point. The proposed analytical approach
makes it possible to control for the first time the thermodynamics of gluon matter at low
temperatures, below $T_c$. We reproduce the properties of the so-called "fuzzy bag"-type models through
the presence of the mass gap in our equation of state. An analytic calculation of the gluon condensate
is obtained as a function of temperature.
\end{abstract}


%\keywords{}

\maketitle

\section{Introduction}

The prediction of a possible existence of the quark-gluon plasma
(QGP) created in the relativistic heavy ion collisions
is one of the most interesting theoretical achievements of Quantum
Chromodynamics (QCD) at non-zero temperatures and densities. A fairly
full list of the relevant pioneering papers is given
in \cite{1,2,3}) and the present status of the
investigations of the properties of QCD matter is described in
\cite{4,5}.

The equation of state (EoS) for the QGP has been
derived analytically up to the order $g^6 \ln(1/g^2)$ by using the
perturbation theory (PT) expansion for the evaluation of the
corresponding thermodynamic potential term by term (\cite{6,7} and
references therein).
However, the most characteristic feature of the thermal PT expansion is
its non-analytical dependence on the coupling constant $g^2$, which means that PT QCD is not
applicable at finite temperatures. The problem is not the poor convergence of this
series \cite{6,7,8,9} but rather the fact that a radius of convergence cannot even be defined;
any next calculated term can be
bigger than the previous one. This is an in-principle problem which
cannot be overcome. From the strictly mathematical point
of view, four-dimensional QCD at non-zero temperatures effectively
becomes a three-dimensional theory. At the same time,
three-dimensional QCD has more severe infrared singularities
\cite{10} and its coupling constant becomes dimensional. It is as a consequence of this that
the dependence becomes non-analytical when using
the dimensionless coupling constant $g^2$. One also needs to introduce
three different scales, $T$, $gT$ and $g^2T$, where $T$ is the
temperature, in order to try somehow to understand the dynamics of the
QGP within the thermal PT QCD approach.

At present, the only practical method to investigate the problem
is lattice QCD at finite temperature and baryon
density, which has recently shown rapid progress
(\cite{4,5,11,12,13,14,15} and references therein). However, lattice QCD,
being a very specific regularization scheme, is primary aimed
at obtaining well-defined corresponding expressions in order
to get realistic numbers for physical quantities. One may therefore get numbers and curves without
understanding what the physics is behind them. Such understanding can
only come from the dynamical theory, which is continuous QCD. For
example, any description of the QGP has to be formulated within the
framework of a dynamical theory. The need for an analytical EoS
remains, but, of course, it should be essentially non-perturbative
(NP), approaching the so-called Stefan-Boltzmann (SB) limit only at
very hight temperatures. Thus the approaches of analytic NP QCD and lattice QCD
to finite-temperature QCD do not exclude each other, but, on the
contrary, should be complementary. This is especially
true at low temperatures where the thermal QCD lattice
calculations suffer from big uncertainties \cite{4,5,11,12,13,14,15}.
On the other hand, any analytic NP approach has to correctly reproduce thermal lattice QCD results at high
temperatures (see papers cited above).
There already exist interesting analytical approaches based on
quasi-particle picture \cite{16,17,18,19,20,21,22,23,24,25} (and references therein) to
analyze results of $SU(3)$ lattice QCD calculations for the QGP EoS.

The main purpose of this paper is to derive the NP analytical EoS for the
gluon matter (GM), i.e., a system consisting purely of Yang-Mills (YM) fields without quark degrees of freedom.
The formalism we use to generalize it to
non-zero temperatures is the effective potential approach for
composite operators \cite{26}. In the absence of external sources
it is nothing but the vacuum energy density (VED). The approach is
NP from the very beginning, since it deals with the
expansion of the corresponding skeleton vacuum loop diagrams.
The key element is the extension of our initial work \cite{27} to non-zero temperatures. This makes it possible to introduce the temperature-dependent bag constant (pressure) as a function of the mass gap. It is this which
is responsible for the large-scale structure of the QCD ground state. The confining dynamics in the GM will therefore be nontrivially taken into account directly through the mass gap and via the temperature-dependent bag constant itself, but other NP effects will be also present. Let us note that the temperature-dependent bag constant
within the thermal PT QCD has been introduced into the Gibbs equilibrium criteria for a phase transition
\cite{28} (see also \cite{16}). The effective potential approach has been already used in order to study the
structure of QCD at very large baryon density for an arbitrary number of flavors \cite{29}.

The present paper is organized as follows. In section 2 the effective potential approach for composite operators
is discussed in general terms. An explicit expression is obtained for the VED for pure YM fields. In section 3 the expression for the gluon pressure is derived taking into account the expression for the bag constant at
zero temperature as a function of the mass gap \cite{27}. Taken together this makes it possible to derive a formula for the pressure that is suitable for the generalization to non-zero temperatures. This consists of the two
independent parts, describing the NP and PT contributions to the gluon pressure. In section 4 the generalization
to non-zero temperatures is performed using the imaginary-time formalism. All the analytic results for the gluon pressure as a function of temperature are collected together in section 5.
In section 6 the expressions are given for the main thermodynamic quantities, such as the entropy and energy densities, the heat capacity, etc., as functions of the pressure. In section 7 we discuss all the numerical results for the NP part of the gluon pressure. Section 8 concludes our discussion. Some explicit expressions for the summation of the thermal logarithms over the Matsubara frequencies are present
in appendix A, while in appendix B a scale-setting scheme of our calculations is formulated.

\section{The Vacuum Energy Density }

\begin{figure}[b]
\begin{center}
\includegraphics[width=0.6\textwidth]{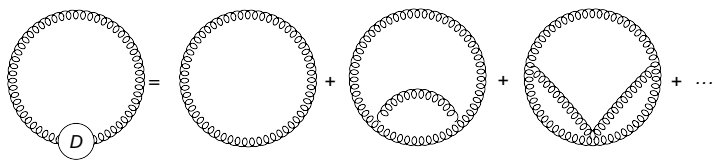}
\caption{Infinite series for the gluon part of the VED to log-loop
level.}
\label{fig:1}

\end{center}
\end{figure}

The quantum part of the VED is determined by the effective
potential approach for composite operators \cite{26}. It is given
in the form of the skeleton loops expansion, containing all the
types of the QCD full propagators and vertices (for its pictorial representation, see \cite{27}).
Each vacuum skeleton loop is itself a sum of an infinite number of the
corresponding PT vacuum loops (i.e., containing the point-like
vertices and free propagators). Thus the effective potential approach makes it possible to calculate
the VED from first principles, i.e., using only fundamental constituents of QCD: gluons and quarks
and their interactions or, more precisely, their propagators and vertices of their interactions.
The number of the vacuum skeleton loops is equal to the power of the Planck constant, $\hbar$, so that
a series expansion for the effective potential is nothing other than the semiclassical WKB loop expansion, which, in general, is an asymptotic series. It has been widely used in quantum field theory \cite{30}. In QCD the instantons have been discovered by using just this method \cite{31} (and references therein).

It is instructive to begin with some general expressions at zero temperature in this formalism.
The gluon part of the VED to leading
order (the so-called log-loop level $\sim \hbar$, whose infinite
series is shown in Fig. 1) is analytically given by the effective
potential for composite operators \cite{26} as follows:

\begin{equation}
V(D) =  { i \over 2} \int {\dd^4q \over (2\pi)^4}
Sp \{ \ln (D_0^{-1}D) - (D_0^{-1}D) + 1 \},
\end{equation}
where $D(q)$ is the full gluon propagator and $D_0(q)$ is its free
counterpart. The traces over space-time and color group indices
are assumed. It is clear from this that the effective potential is normalized to
the free PT vacuum to be zero, i.e., $V(D_0) = 0$. We note that the YM bag constant has been
calculated to this order in \cite{27}.

The two-point Green's function, describing the full gluon
propagator, is

\begin{equation}
D_{\mu\nu}(q) = - i \left\{ T_{\mu\nu}(q)d(-q^2; \xi) + \xi
L_{\mu\nu}(q) \right\} {1 \over q^2 },
\end{equation}
where $\xi$ is the gauge-fixing parameter and

\begin{equation}
T_{\mu\nu}(q) = g_{\mu\nu} - {q_\mu q_\nu \over q^2} = g_{\mu\nu }
- L_{\mu\nu}(q).
\end{equation}

Its free counterpart $D_0 \equiv D^0_{\mu\nu}(q)$ is obtaining by
replacing the full gluon propagator's Lorentz structure $d(-q^2; \xi)$ in Eq.~(2) by unity, i.e.,

\begin{equation}
D^0_{\mu\nu}(q) = - i \left\{ T_{\mu\nu}(q) + \xi L_{\mu\nu}(q)
\right\} {1 \over q^2}.
\end{equation}

In order to evaluate the effective potential (1) with
Eq.~(2), we use the well-known identity \cite{32}

\begin{equation}
Sp \ln (D_0^{-1}D) = 8 \times 4 \ln det (D_0^{-1}D) = 32 \ln [ (3/
4 )d(-q^2; \xi) + (1 / 4 ) ].
\end{equation}

 Going over to four-dimensional Euclidean space in Eq.~(1),
one obtains ($\epsilon_g = V(D)$)

\begin{equation}
\epsilon_g = - 16 \int {\dd^4q \over (2\pi)^4} \left[ \ln [1 + 3
d(q^2; \xi)] - {3 \over 4}d(q^2; \xi) + a \right],
\end{equation}
where the constant $a = (3/4) - 2 \ln 2$ and an
integration from zero to infinity over $q^2$ is assumed. The VED
$\epsilon_g$ derived in Eq.~(6) is already a colorless (color-singlet) quantity,
since it has been summed over the color indices. It also does not
depend explicitly on the unphysical (longitudinal) part of the
full gluon propagator due to the product $(D_0^{-1}D)$ which, in turn, comes from the above-mentioned
normalization of the free PT vacuum to zero.
Thus it is worth emphasizing that only the transversal ("physical")
degrees of freedom only of the gauge bosons contribute to this
equation. As a consequence, in the effective potential approach to leading order there is no need
for the ghost degrees of freedom from the very beginning in order
to cancel the longitudinal ("unphysical") component of the full
gluon propagator. This role is played by the normalization condition. However, this does not work for the higher
order vacuum skeleton loops. In this case and beyond the PT at any gauge (i.e., in the general case) the cancelation of unphysical gluon modes should proceed with the help of ghosts \cite{27}.

An overall numerical factor $1/2$ has been introduced
into Eq.~(1) in order to make the gluon degrees of freedom
equal to $32/2 = 16 = 8 \times 2$, where $8$ color of gluons times $2$
helicity (transversal) degrees of freedom (see Eqs.~(5) and (6)).

In what follows the Lorentz structure $d(q^2) \equiv d(q^2; \xi)$ will be
called the full effective charge ("running") or the gluon propagator invariant function (its form factor), for convenience. For the generalization of Eq.~(6) to non-zero
temperatures, the most important thing is to introduce correctly the above-mentioned bag constant.

\section{The gluon pressure at zero temperature}

The vacuum of QCD is a very complicated confining medium and its
dynamical and topological complexity means that its structure can
be organized at various levels: classical and quantum (see, for
example \cite{33,34,35} and references therein). It is mainly
NP by origin, character and magnitude, since the corresponding
fine structure constant is large. However, the virtual gluon
field configurations and excitations of the PT magnitude due to
asymptotic freedom (AF) \cite{36} are also present there.

One of the main dynamical characteristics of the true QCD ground
state is the bag constant $B$. Its name has come from the famous bag
models for hadrons \cite{37,38}, but its present understanding (and thus modern
definition) has nothing to do with the hadron properties. It is
defined as the difference between the PT and the NP VEDs \cite{39,40,41,42}.
We can symbolically write $B = VED^{PT} - VED$, where the $VED$ is
NP but "contaminated" by the PT contributions (i.e., this is a
full $VED$ like a full gluon propagator). Rewriting $B$ as
$B = VED^{PT} - VED = VED^{PT} - [VED -
VED^{PT} + VED^{PT}] = VED^{PT} - [VED^{TNP} + VED^{PT}] = -
VED^{TNP} > 0$, since the VED is always negative. The bag constant
is nothing but the truly NP (TNP) VED, apart from the sign, by
definition, and thus is free of the PT contributions
("contaminations"). In order to consider it also as a physical
characteristic of the true QCD ground state, the bag constant
correctly calculated should satisfy some other
requirements such as colorlessness and dependence on the physical
degrees of freedom, finiteness, gauge-independence, no imaginary
part (stable vacuum), etc.

In our previous work \cite{27}, we have already derived an expression for the bag constant, which
satisfies the above conditions:

\begin{equation}
B_{YM} = 16 \int^{q^2_{eff}} {\dd^4q \over (2\pi)^4} \left[ \ln [1
+ 3 d^{TNP}(q^2)] - {3 \over 4} d^{TNP}(q^2) \right],
\end{equation}
where symbolically shown $q^2_{eff}$ is the effective scale
squared, separating the soft momenta from the hard ones in the
integration over $q^2$. This is a general expression for any TNP
effective charge in order to calculate the bag constant from first principles. It is
defined as the special function of the TNP effective charge
integrated out over the NP region (soft momenta region, $0 \leq
q^2 \leq q^2_{eff}$).

Adding the bag constant to the both sides of Eq.~(6), and introducing the gluon pressure
$P_g = \epsilon_g + B_{YM}$, one obtains

\begin{equation}
P_g = B_{YM} - 16 \int {\dd^4q \over (2\pi)^4} \left[ \ln [1 + 3
d(q^2)] - {3 \over 4}d(q^2) + a \right].
\end{equation}

The next step is to establish the relation between the full effective charge $d(q^2)$ and its
TNP counterpart $d^{TNP}(q^2)$. In our previous work \cite{43} it has been proven that
the full effective charge depends explicitly and regularly on the scale $\Delta^2$
(the so-called mass gap) responsible for the NP dynamics in QCD, i.e., $d(q^2) \equiv (q^2; \Delta^2)$.
The above-mentioned symbolic subtraction at the fundamental gluon level can then be defined as \cite{27,43}:
$d^{TNP}(q^2; \Delta^2) = d(q^2; \Delta^2) - d(q^2; \Delta^2=0) =
d(q^2; \Delta^2)  - d^{PT}(q^2)$.
In this way the separation between the TNP effective charge and its PT counterpart becomes
exact, but not unique. On how to make this separation exact and unique
at the same time see subsection below. Evidently, both
the TNP part and its PT counterpart are valid throughout the whole energy/momentum range, i.e, they are
not asymptotics. Let us also emphasize the principle difference
between $d(q^2)$ and $d^{TNP}(q^2)$. The former is the NP
quantity "contaminated" by PT contributions, while the latter
one, being also NP, is free of them.

However, this is not the whole story yet. Since the PT part $d^{PT}(q^2)$ contains the free gluon Lorentz
structure $d^0(q^2)=1$, it should be extracted explicitly as follows, where we have dropped the explicit
dependence on $\Delta^2$:

\begin{equation}
d(q^2) = d^{TNP}(q^2) + d^{PT}(q^2)=
d^{TNP}(q^2) + 1 + d^{AF}(q^2),
\end{equation}
and $d^{AF}(q^2)$ describes the part responsible for AF in the
PT gluon effective charge. This procedure is necessary in order to
maintain the normalization of the free PT vacuum to
zero. In this connection, let us stress that the second equality in this relation does not
imply any violation of AF in the full gluon effective charge $d(q^2)$, provided by the PT effective charge
$d^{PT}(q^2)$ in the first equality of the same relation. Extracting $d^0=1$ explicitly, we thereby subtract the divergent contribution associated with the constant $a$ in Eq.~(6), and thus the above-mentioned normalization condition will be automatically satisfied. We now replace all the effective
charges as follows: $d^{TNP}(q^2) \equiv \alpha_s^{TNP}(q^2)$ and
$d^{AF}(q^2) \equiv \alpha^{AF}(q^2)$. The explicit expression for the
effective charge responsible for AF, that's $\alpha^{AF}(q^2)$,  will be given
in part II of our work, since its explicit expression is not used here.

Substituting the decomposition (9) into Eq.~(8), lengthy algebra leads to

\begin{equation}
P_g = P_{NP} + P_{PT}= B_{YM} + P_{YM} + P_{PT},
\end{equation}
i.e., $P_{NP} = B_{YM} + P_{YM}$. In Eq.~(10) $B_{YM}$ is given in Eq.~(7), while

\begin{equation}
P_{YM} = - 16 \int {\dd^4q \over (2\pi)^4} \left[ \ln [1 + {3
\over 4} \alpha_s^{TNP}(q^2)] - {3 \over 4} \alpha_s^{TNP}(q^2)
\right],
\end{equation}
and

\begin{equation}
P_{PT}= - 16  \int {\dd^4q \over (2\pi)^4} \left[ \ln [1 + {3
\alpha^{AF}(q^2) \over 4 + 3 \alpha_s^{TNP}(q^2)}] - {3 \over 4}
\alpha^{AF}(q^2) \right].
\end{equation}
$P_{YM}$ in Eq.~(11) depends exclusively on the TNP effective
charge. Together with the bag constant (7) it forms the NP part of the gluon pressure (10).
$P_{PT}$ in Eq.~(12) contains the contribution which is mainly determined by
the AF part of the PT effective charge, though the dependence on
the TNP effective charge is also present (it is logarithmically
suppressed in comparison to the pure AF term). If the interaction is switched
off, i.e., putting formally $\alpha^{AF}(q^2)= \alpha_s^{TNP}(q^2)=0$,
then $P_g=0$ in accordance with the initial normalization of the free PT vacuum to zero. Evidently, just the right-hand-side of Eq.~(10) is to be generalized to non-zero temperature, on account of the explicit expressions (7), (11) and (12).

\subsection{Confining effective charge}

The only problem remaining is the explicit expression for the TNP effective charge. Evidently, it
has to be consistent with our initial work \cite{27}, where the bag constant was evaluated at zero temperature,
i.e.,

\begin{equation}
\alpha_s^{TNP}(q^2) \rightarrow \alpha_s^{INP}(q^2) = {\Delta^2 \over q^2},
\end{equation}
where the superscript "INP" stands for the intrinsically NP effective charge. Here $\Delta^2 \equiv \Delta^2_{JW}$ is the Jaffe-Witten (JW) mass gap, mentioned above, which is responsible for the large-scale structure of the QCD vacuum, and thus for its INP dynamics \cite{44}. Let us note that how the mass gap appears in QCD has been shown in our work \cite{43}.

A few additional remarks are in order.
In \cite{45} it has been shown that the TNP part of the full gluon propagator as a function of the mass gap also contains a term that is regular at origin. It is for this reason that it is not uniquely
separated from the PT gluon propagator, which effective charge that is always regular at origin.
We distinguish between the INP and the PT effective charges not only by the presence of the mass gap,
but by the character of the IR singularities as well \cite{45,46}. So only after the replacement (13) the
INP effective charge is uniquely and exactly separated from its PT counterpart, and
the obtained expression for the bag constant (7) becomes now {\it free of all the types of the PT contributions ("contaminations")}.

In \cite{46} we have shown that the so-called INP gluon propagator is, by construction, purely transversal in a gauge invariant way. It exactly converges to the gluon propagator, whose effective charge is
Eq.~(13), after the renormalization programme of the regularized mass gap is performed. This result has been
obtained in the most general way, i.e.,
without making any truncations/approximations/assumptions or choosing a special gauge. {\it Hence the expression (13) is not an ansatz, and thus it is mathematically well justified}. However, it should also be physically well justified. The problem is that the gluon equation of motion is highly nonlinear, so the number of independent exact solutions is not fixed $a \ priori$ (this number may be even bigger, depending on the different truncations/approximations/assumptions and the concrete gauge choice made). They should be considered on equal footing from the very beginning. In our previous work \cite{45}, at least the two different general types of exact solutions for the full gluon propagator have been found as a function of the
regularized mass gap: the first is smooth at small gluon momentum, allowing for the gluons to acquire an effective gluon mass (the so-called massive solution). The second is singular in the $q^2 \rightarrow 0$ limit, so that the gluons always remain massless (the so-called nonlinear iteration solution). Only this solution has an affective charge (13) after the renormalization is completed. For the detailed arguments which makes its choice justified from the physical point of view as well (i.e., emphasizing its confining nature) see our paper \cite{27}. In addition, let us note that the above-mentioned massive solution is not confining.

\section{Generalization to non-zero temperatures}

In the imaginary-time formalism \cite{7,47}, all the
four-dimensional integrals can be easily generalized to non-zero
temperatures $T$ according to the prescription (note that
there is already Euclidean signature)

\begin{equation}
\int {\dd q_0 \over (2\pi)} \rightarrow T \sum_{n=- \infty}^{+
\infty}, \quad \ q^2 = {\bf q}^2 + q^2_0 = {\bf q}^2 +
\omega^2_n = \omega^2 + \omega^2_n, \ \omega_n = 2n \pi T,
\end{equation}
i.e., each integral over $q_0$ of a loop momentum is to be
replaced by the sum over the Matsubara frequencies labeled by
$n$, which obviously assumes the replacement $q_0 \rightarrow
\omega_n= 2n \pi T$ for bosons (gluons). In frequency-momentum
space the INP effective charges (13) becomes

\begin{equation}
\alpha_s^{INP}(q^2) = \alpha^{INP}_s({\bf q}^2, \omega_n^2) =
\alpha^{TNP}_s( \omega^2, \omega_n^2) = { \Delta^2 \over \omega^2
+ \omega_n^2},
\end{equation}
and it is also convenient to introduce the following notations:

\begin{equation}
T^{-1} = \beta, \quad \omega = \sqrt{{\bf q}^2},
\end{equation}
$\alpha^{AF}(q^2) = \alpha^{AF}({\bf q}^2, \omega_n^2) =
\alpha^{AF} (\omega^2, \omega_n^2)$,
where, evidently, in all the expressions here and below ${\bf q}^2$
is the square of the three-dimensional loop momentum, in complete agreement
with the relations (14).

Introducing the temperature dependence into the right-hand-side of
the relation (10), we finally obtain

\begin{equation}
P_g(T) = P_{NP}(T) + P_{PT}(T) = B_{YM}(T) + P_{YM}(T) + P_{PT}(T).
\end{equation}

\subsection{Derivation of $B_{YM}(T)$}

It is convenient to begin with Eq.~(7) for the bag constant. In frequency-momentum space
the temperature-dependent YM bag pressure becomes

\begin{equation}
B_{YM}(T) =  16 \int {\dd^3q \over (2\pi)^3} \ T \sum_{n= -
\infty}^{+ \infty} \left[ \ln [1 + 3 \alpha_s^{TNP}({\bf q}^2,
\omega^2_n)] - {3 \over 4} \alpha_s^{TNP}({\bf q}^2, \omega^2_n)
\right].
\end{equation}
After the substitution of the expression from Eq.~(15), one obtains

\begin{equation}
B_{YM}(T) = 16 \int {\dd^3q \over (2\pi)^3} \ T \sum_{n= -
\infty}^{+ \infty} \left[ \ln \left( {\omega'^2 + \omega^2_n \over
\omega^2 + \omega^2_n} \right) - {3 \over 4} { \Delta^2 \over \omega^2 +
\omega^2_n} \right],
\end{equation}
where we introduced the notations:

\begin{equation}
\omega' = \sqrt{{\bf q}^2 + 3 \Delta^2} = \sqrt{\omega^2 +
m'^2_{eff}}, \quad m'_{eff}= \sqrt{3} \Delta.
\end{equation}

One of the attractive features of the confining effective charge (15) is
that it allows for an exact summation over the Matsubara
frequencies (see appendix A). Substituting all our results of the summation into Eq. (19), dropping
the $\beta$-independent terms \cite{7}, and performing almost trivial integration over angular
variables, one gets

\begin{equation}
B_{YM}(T) =  - {8 \over \pi^2} \int_0^{\omega_{eff}} \dd\omega \ \omega^2 \left[ {3
\over 4} \Delta^2 {1 \over \omega} { 1 \over e^{\beta \omega} - 1
} - 2 \beta^{-1} \ln \left( {1 - e^{- \beta \omega'} \over 1 -
e^{- \beta \omega}} \right) \right].
\end{equation}
The $\omega_{eff}$, which is the three-dimensional analog of $q_{eff}$ in Eq.~(7), is discussed in appendix B, where numerical values are given.

It is convenient to present the integral (21) as a sum of several terms

\begin{equation}
B_{YM}(T) = - {6 \over \pi^2} \Delta^2 B_{YM}^{(1)}(T) - {16 \over
\pi^2} T \left[ B_{YM}^{(2)}(T) - B_{YM}^{(3)}(T) \right],
\end{equation}
where the explicit expressions of all the terms are given as

\begin{eqnarray}
B_{YM}^{(1)}(T) &=& \int_0^{\omega_{eff}} \dd\omega {\omega \over
e^{\beta\omega} -1}, \nonumber\\
B_{YM}^{(2)}(T)  &=& \int_0^{\omega_{eff}} \dd\omega \ \omega^2 \ln
\left( 1- e^{-\beta \omega} \right), \nonumber\\
B_{YM}^{(3)}(T)  &=& \int_0^{\omega_{eff}} \dd\omega \ \omega^2 \ln
\left( 1- e^{-\beta \omega'} \right).
\end{eqnarray}
Here and below $N=(e^{\beta\omega} -1)^{-1}$ is the Bose-Einstein distribution (the so-called gluon mean number  \cite{7}).

\begin{figure}
\begin{center}
\includegraphics[width=10cm]{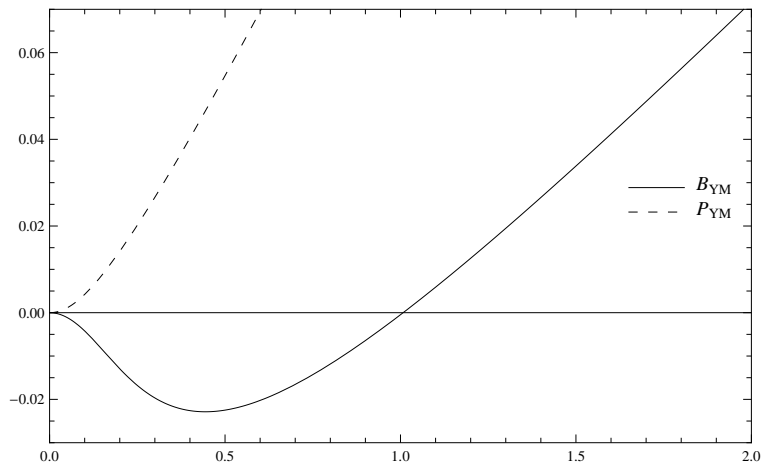}
\caption{The bag constant (22) and the NP YM part (27) in $\GeV^4$ units as functions of temperature $T$ in $\GeV$ units. The bag constant at $T = 1 \ \GeV$ is zero, so up to this value it is responsible
for the NP vacuum contributions to the pressure (17) and hence (29).}
\label{fig:2}
\end{center}
\end{figure}

\subsection{Derivation of $P_{YM}(T)$}

In frequency-momentum space Eq.~(11) becomes

\begin{equation}
P_{YM}(T) = - 16 \int {\dd^3q \over (2\pi)^3} \ T \sum_{n= -
\infty}^{+ \infty} \left[ \ln [1 + {3 \over 4} \alpha_s^{TNP}({\bf
q}^2, \omega^2_n)]  - {3 \over 4} \alpha_s^{TNP} ({\bf q}^2,
\omega^2_n) \right],
\end{equation}
thus determining the temperature-dependent YM part of the NP pressure of Eq.~(17).
After substituting of the expression (15), one obtains

\begin{equation}
P_{YM}(T) = - 16 \int {\dd^3q \over (2\pi)^3} \ T \sum_{n= -
\infty}^{+ \infty} \left[ \ln \left( {\bar \omega^2 + \omega^2_n \over
 \omega^2 + \omega^2_n} \right) - {3 \over 4} { \Delta^2  \over \omega^2
+ \omega^2_n} \right],
\end{equation}
where we introduced the following notations:

\begin{equation}
\bar \omega = \sqrt{{\bf q}^2 + {3 \over 4} \Delta^2} =
\sqrt{\omega^2 + \bar m^2_{eff}}, \quad \bar m_{eff}= {\sqrt{3}
\over 2} \Delta = {1 \over 2} m'_{eff}.
\end{equation}

Comparing Eqs.~(19) and (25) one can write down the final result
directly. For this purpose, in the system of Eqs. (22)-(23)
one must change the overall sign, replace $\omega'$ by $\bar
\omega$ and integrate from zero to infinity. Thus, one obtains

\begin{equation}
P_{YM}(T) = {6 \over \pi^2} \Delta^2 P_{YM}^{(1)}(T) + {16 \over
\pi^2} T \left[ P_{YM}^{(2)}(T) - P_{YM}^{(3)}(T) \right],
\end{equation}
where

\begin{eqnarray}
P_{YM}^{(1)}(T) &=& \int_0^{\infty} \dd\omega {\omega \over
e^{\beta\omega} -1} = {\pi^2 \over 6} T^2, \nonumber\\
P_{YM}^{(2)}(T)  &=& \int_0^{\infty} \dd\omega \ \omega^2 \ln \left(
1- e^{-\beta \omega} \right), \nonumber\\
P_{YM}^{(3)}(T)  &=& \int_0^{\infty} \dd\omega \ \omega^2 \ln \left(
1- e^{-\beta \bar \omega} \right).
\end{eqnarray}
The bag pressure (22) and the NP YM part (27) in $\GeV^4$ units as functions of temperature $T$
are shown in Fig. 2. The scale-setting scheme of all our numerical calculations in this paper is presented
in appendix B.

\section{The gluon pressure at non-zero temperatures}

Summing up all the expressions and integrals (22)-(23) and
(27)-(28), the gluon pressure (17) becomes

\begin{equation}
P_g(T) = P_{NP}(T) + P_{PT}(T),
\end{equation}
where

\begin{eqnarray}
P_{NP}(T) &=& B_{YM}(T) + P_{YM}(T) \nonumber\\
&=& {6 \over \pi^2} \Delta^2 P_1 (T) + {16 \over \pi^2} T
[P_2(T) + P_3(T) - P_4(T)],
\end{eqnarray}
and

\begin{equation}
P_1(T) = \int_{\omega_{eff}}^{\infty} \dd \omega {\omega \over
e^{\beta\omega} -1},
\end{equation}
while

\begin{eqnarray}
P_2(T) &=& \int_{\omega_{eff}}^{\infty} \dd \omega \ \omega^2
\ln \left( 1- e^{-\beta\omega} \right), \nonumber\\
P_3(T)&=& \int_0^{\omega_{eff}} \dd \omega \ \omega^2 \ln
\left( 1 - e^{- \beta\omega'} \right), \nonumber\\
P_4(T) &=& \int_0^{\infty} \dd \omega \ \omega^2 \ln \left( 1 -
e^{- \beta \bar \omega} \right).
\end{eqnarray}
Let us recall once more that in all these integrals $\beta =
T^{-1}$, $\omega_{eff}$ along with the mass gap $\Delta^2$ is
fixed (appendix B), while $\omega'$ and $\bar \omega$ are
given by the relations (20) and (26), respectively. In the
formal PT $\Delta^2=0$ limit it follows that $\bar \omega =
\omega' = \omega$ and the combination $P_2(T) + P_3(T) - P_4(T)$
becomes identically zero. Thus the NP part (30) of the gluon pressure (29)
in this limit vanishes.

In frequency-momentum space the PT part (12) of the gluon pressure (29) is

\begin{equation}
P_{PT}(T) = - { 8 \over \pi^2}  \int_0^{\infty} \dd\omega \ \omega^2 \ T
\sum_{n= - \infty}^{+ \infty} \left[ \ln \left( 1 + {3
\alpha^{AF}(\omega^2, \omega^2_n) \over 4 + 3
\alpha_s^{TNP}(\omega^2, \omega^2_n)} \right) - {3 \over 4}
\alpha^{AF}(\omega^2, \omega^2_n) \right],
\end{equation}
where the trivial integration over angular variables has already been carried out.
Unfortunately, one cannot perform analytically (i.e., exactly) the summation over the
Matsubara frequencies in this integral, using the AF expression for
$\alpha^{AF}(\omega^2, \omega^2_n)$ \cite{36,46}. At this stage the PT
part (33) remains undetermined and will
be numerically evaluated elsewhere. Note that, since it becomes zero when the interaction
is switched formally off ($\alpha^{AF}(\omega^2, \omega^2_n)=0$), both terms (30) and (33)
and hence the gluon pressure (29) satisfy the initial normalization condition.

Confining dynamics in Eq.~(29) is implemented via the
bag pressure (22) and the mass gap itself. However, other NP effects are also present via $P_{YM}(T)$
and $P_{PT}(T)$, given in Eqs.~(27) and (33), respectively, though in Eq.~(33) they are logarithmically suppressed.

\section{Main thermodynamic quantities}

Together with the pressure $P(T)$, the main thermodynamic
quantities are the entropy density $s(T)$ and the energy density
$\epsilon(T)$. The general formulae which connect them are
\cite{7}

\begin{eqnarray}
s(T) &=& {\partial P(T) \over \partial T}, \nonumber\\
\epsilon(T) &=&  T \left( \partial P(T) \over \partial T \right) -
P(T)= T s(T) - P(T)
\end{eqnarray}
for pure YM fields, i.e., when the chemical potential vanishes.
Let us note that in quantum statistics the pressure
$P(T)$ is up to a sign equal to the thermodynamic potential
$\Omega(T)$, i.e., $P(T) = - \Omega(T) > 0$ \cite{7}.

Other thermodynamic quantities of interest are the heat capacity
$c_V(T)$ and the velocity of sound squared $c^2_s(T)$, which are defined
as follows:

\begin{equation}
c_V(T) = { \partial \epsilon(T) \over \partial T}  = T \left(
\partial s(T) \over \partial T \right),
\end{equation}
and

\begin{equation}
c_s^2(T) = { \partial P(T) \over \partial \epsilon(T)}  = { s(T)
\over c_V(T)},
\end{equation}
i.e., they are defined through the second derivative of the pressure. The so-called conformity

\begin{equation}
C(T) = { P(T) \over \epsilon (T)}
\end{equation}
mimics the behavior of the speed of sound squared (36) but without involving such a differentiation.

The thermodynamic quantity of a special interest is the thermal expectation value of the trace of the energy momentum tensor. This trace anomaly relation measures the deviation of the difference

\begin{equation}
\epsilon(T) - 3P(T)
\end{equation}
from zero at finite temperatures, in the high temperature limit it must vanish. As a consequence
it is very sensitive to the NP contributions to the EoS. It also assists in the temperature dependence of the gluon condensate \cite{48} (see \cite{49}), namely

\begin{equation}
<G^2>_T = <G^2>_0 - [ \epsilon(T) - 3P(T)],
\end{equation}
where $<G^2>_0 \equiv <G^2>_{T=0}$ denotes the gluon condensate at zero temperature, whose
numerical value is discussed in appendix B.

As mentioned above, the main purpose of this paper is to investigate analytically and calculate the NP part (30) of Eq.~(29), while the corresponding investigation of the PT part (33) will be the subject of subsequent work.
This will make it possible to complete the investigation, derivation and final calculation of the full GM EoS.

\subsection{SB limit}

The high-temperature behavior of all the thermodynamic quantities
is governed by the SB ideal gas limit, when the
matter can be described in terms of non-interacting massless
particles (gluons). In this limit these quantities satisfy
special relations \cite{7}

\begin{equation}
{3P_{SB}(T) \over T^4} = {\epsilon_{SB}(T) \over T^4} = {3
s_{SB}(T) \over 4 T^3} = {c_{V(SB)}(T) \over 4 T^3} = { 24 \over
45} \pi^2 \approx 5.26, \ T \rightarrow \infty,
\end{equation}
and

\begin{equation}
C_{SB}(T) = c_{s(SB)}^2(T) = { 1 \over 3}, \quad T \rightarrow \infty,
\end{equation}
on account of the previous relations and their definitions in Eqs.~(36) and (37).
The trace anomaly relation (38) also satisfies the SB limit, namely

\begin{equation}
\epsilon_{SB}(T)  - 3 P_{SB}(T) = 0, \quad T \rightarrow \infty,
\end{equation}
as it comes out from the relations (40). In what follows its right-hand-side
will be called the general SB number.

\begin{figure}
\begin{center}
\includegraphics[width=10cm]{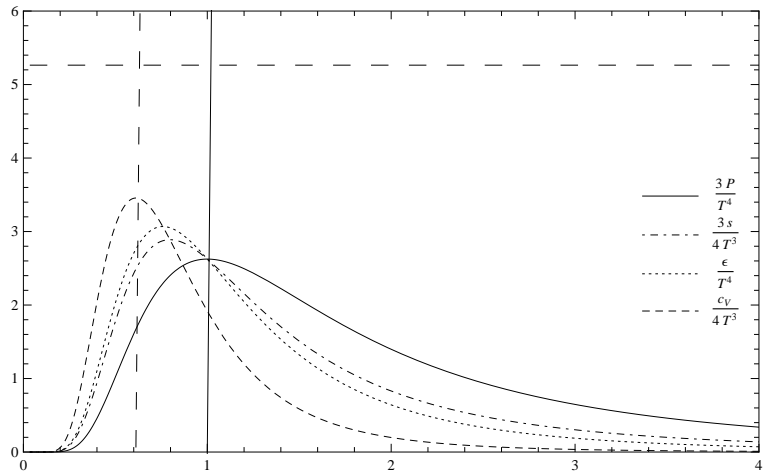}
\caption{The NP pressure, the entropy and energy densities, the heat
capacity as functions of $T/T_c$. The NP pressure has a maximum
at $T_c=266.5 \ \MeV$. Here the horizontal dashed
line is the general SB number (40), while the vertical dashed
line at $0.6T_c$ separates the low-temperatures region from the
transition region $(0.6 - 1)T_c$.}
\label{fig:3}
\end{center}
\end{figure}

\begin{figure}
\begin{center}
\includegraphics[width=10cm]{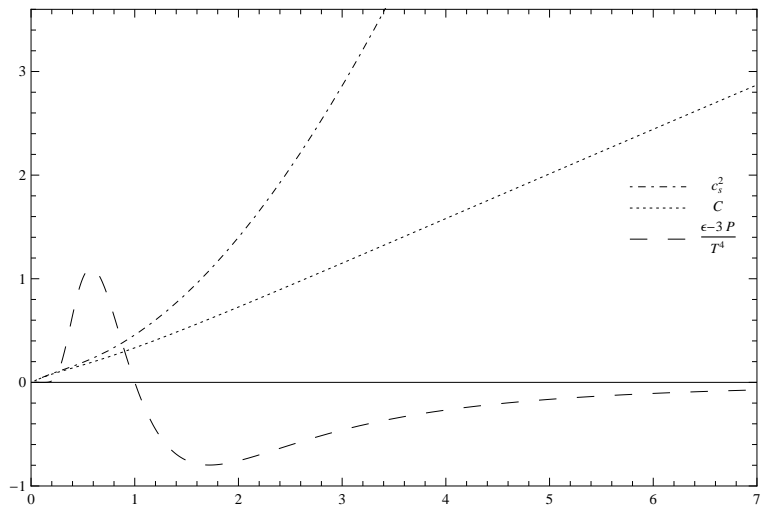}
\caption{The NP velocity of sound, conformity and the trace anomaly relation as
functions of $T/T_c$.}
\label{fig:4}
\end{center}
\end{figure}

\begin{figure}
\begin{center}
\includegraphics[width=10cm]{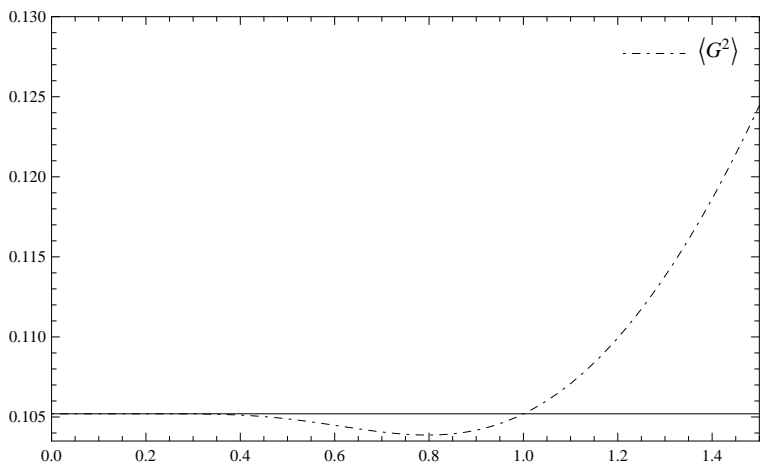}
\caption{The NP gluon condensate (39) in $\GeV^4$ units as a function of $T/T_c$. It shows
little temperature dependence below $T_c$ and grows rapidly above $T_c$.
Here the solid line is its value at zero temperature $<G^2>_0$ (see appendix B).}
\label{fig:5}
\end{center}
\end{figure}

\section{Numerical results and discussion }

All our numerical results obtained for the thermodynamical quantities discussed in the previous section
and calculated with the help of the NP pressure (30) are shown in Figs. 3, 4 and 5.
The NP gluon pressure has a maximum at some finite ("characteristic") temperature, $T=T_c = 266.5 \ \MeV$.
This means that it may change continuously
its regime in the near neighborhood of this point in order for its full counterpart to
achieve the thermodynamic SB limit (40) at high temperatures. However, this is not possible
for the other NP thermodynamic quantities, as is seen from
the curves shown in Fig. 3 (none of the power-type fall off around $T_c$
can be continuously transformed into the almost constant
behavior at high temperatures). In order to achieve the corresponding
thermodynamic SB limits at high temperatures their full
counterparts would have to undergo drastic changes around this point.
As we already know from the thermodynamics of $SU(3)$ lattice QCD \cite{49} (see
\cite{50,51,52}) the energy and entropy densities have
discontinuities (jump or, equivalently, step discontinuities) at a temperature of about $T_c=260 - 270 \ \MeV$. Our
characteristic temperature $T_c=266.5 \ \MeV$ is surprisingly very
close to this value. If we were not aware of the thermal lattice QCD results then we would be able to predict them.
Since we are aware of them, these lattice results confirm our expectation of a
sharp changes in the behavior of the entropy and energy densities
in the region where the pressure is continuous.

One of the interesting features of our calculations is seen in Fig. 3.
At the characteristic temperature $T_c$
the NP pressure, the entropy and energy densities satisfy a SB-type relations, namely

\begin{equation}
{3P_{NP}(T_c) \over T^4_c} = {\epsilon_{NP}(T_c) \over T^4_c} = {3
s_{NP}(T_c) \over 4 T^3_c} = { 12 \over 45} \pi^2 \approx 2.63,
\end{equation}
where, obviously, the right-hand-side is
half of the general SB number (40). In turn this yields

\begin{equation}
C_{NP}(T_c) = { 1 \over 3}, \quad  \epsilon_{NP}(T_c)  - 3 P_{NP}(T_c) = 0,
\end{equation}
so that the NP trace anomaly relation approaches zero from below in the $T \rightarrow \infty \ (\beta \rightarrow 0$)
limit (see Fig. 4), i.e, satisfying the SB relation (42) in this limit as well.

Since the NP entropy and energy densities satisfy the SB-type relations (43), we expect for their full counterparts to have step discontinuities at $T_c$, as discussed above.
The NP heat capacity and the velocity of sound squared do not satisfy the
SB-type relations (43), as can be seen in Figs. 3 and 4. We expect, therefore, for the full heat capacity to have an essential discontinuity at $T_c$, since, in general, $[c_V(T_c)]^{-1} =0$, and hence $c_s^2(T_c) =0$ due to the relation (36). They are defined through derivatives of the entropy and energy
densities in Eqs.~(35)-(36), i.e., they involve the second derivatives of the pressure.
That is a reason why these thermodynamic quantities
are too sensitive to the dynamical structure of the GM in the near neighborhood of $T_c$.

Moreover, due to the SB-type relations (43)-(44), and on account of the SB relations (40), for
example it follows that

\begin{equation}
\Bigl[ P_{SB}(T) - 2P_{NP}(T) \Bigr]_{T=T_c} = 0, \
\Bigl \{ {\partial \over \partial T} \Bigl[ P_{SB}(T) - 2P_{NP}(T) \Bigr] \Bigr\}_{T=T_c} = 0.
\end{equation}
Combining these relations again with the SB-type relations (43)-(44) it is possible to derive some
other exact relations between different thermodynamic quantities at $T_c$.
Their importance will be explicitly shown in part II of our investigation, where
they will be extensively exploited.

From Fig. 3 it clearly follows that the temperature range for the NP thermodynamic quantities
can be divided into the three different intervals:

{\it (i). The low-temperatures interval up to its upper bound of about $0.6T_c$}.
It is defined by the first maximum of the heat capacity which
appears in this region before temperature reaches $T_c$. This region is obviously dominated by the NP
contributions (30) to the gluon pressure (29). The behavior of all the NP thermodynamic quantities
are shown in Figs. 5, 6 and 7. We do not expect any serious changes in the
behavior of their full counterparts in this region. However, whatever changes may occur they will
be under our control, since the NP part (30) is exactly
calculated. The important observation is that all the NP thermodynamic quantities which have been properly scaled
go down exponentially as temperature approaches zero. This is a general feature of the behavior of all the full thermodynamic quantities far below $T_c$; otherwise their zero temperature limit would not be realized.
The almost constant behavior of the NP velocity of sound squared and conformity seen in Fig. 7 is due to the fact that they are the ratios of the corresponding thermodynamic quantities, see Eqs.~(36)-(37).
It seems that for the first time it is possible to predict the behavior of all the important thermodynamic quantities
in this interval. We shall refer to the GM in this region as a "confining phase"
(there are no lattice data for this region at all).

{\it (ii). The transition interval $(0.6 - 1)T_c$}.
In this region all the thermodynamic quantities should undergo sharp changes in their
regimes, so that their full counterparts will be able to
achieve their corresponding SB limits at high temperatures. The full
GM pressure remains continuous in this region.
However, the existence of maxima in the behavior of the NP
thermodynamic quantities is expected in this interval, since they should "die" at
high temperatures. The NP pressure is a part of the full GM
pressure, which should be a continuously growing function of $T$ throughout the whole
temperature range. All other full thermodynamic quantities are continuous functions of $T$ only below and above $T_c$. Within our approach they are expected to have discontinuities of different character at $T_c$,
as mentioned above (apart from the gluon condensate). Following the authors of \cite{53}, we shall refer to the GM in the transition region as a "mixed phase". The changes in the dynamical structure of the GM just in this
region will determine the nature of the phase transition at $T_c$.

{\it (iii). The temperatures interval starting at $T_c$}. This interval itself can be clearly subdivided
into two intervals, since in the integrals (31)-(32) $\omega_{eff} = 3.75T_c = 1 \ \GeV$:
{\it (a) The moderate-temperatures $(1-3.75)T_c$}, when the NP effects are still significant and
{\it (b) The high-temperatures starting at $3.75T_c$}, when the NP effects become small.

Therefore, even above $T_c$, the GM can be understood as being in the two different forms. The first
one can be considered as strongly coupled GM, where the NP effects
are still important. The second one can be considered as weakly
coupled GM, where the NP effects become already negligible. The typical temperature which is of
about $3.75T_c = 1 \ \GeV$ for the GM, is too high to be reached even at LHC. For QCD matter this
temperature should be substantially decreased, and therefore be accessible at RHIC, and especially at LHC.
The general feature for the behavior of all the properly scaled NP thermodynamic quantities above $T_c$
is their power-type fall off, while their full counterparts should show a power-type rise at high temperatures.
Apparently, we may refer to the GM in the moderate temperature region as
an "extended mixed phase". The behavior of all the full thermodynamic quantities in both mixed phases is
governed by the NP pressure (30) and by the first term in the PT pressure (33). We shall refer to the
GM in the high temperatures region as an "AF phase", since the behavior of all the full thermodynamic
quantities will be determined by the second term in the PT pressure (33) and the free gluons contribution.

A few remarks in advance are here in order. From our discussion it clearly follows that the approximation of the gluon pressure (29) by its NP part (30) only (though exactly calculated in the present investigation) breaks down at $T_c$, when the NP pressure achieves its maximum. For the derivatives of the pressure, this approximation breaks down even earlier, starting from about $0.6T_c$, as discussed above.
This explains the sign change in the interaction measure and the supersonic value of the speed of sound
squared both at $T_c$ as seen in Fig. 4. The rapid rise of conformity and the speed of sound squared above $T_c$ in Fig. 4 is also unphysical. This also explains why the gluon condensate in Fig. 5 increases drastically above $T_c$. Taking into account the NP (30), PT (33), and free gluons contributions to the full GM pressure, all these unphysical effects will disappear in the full thermodynamic quantities, thus allowing them to achieve their corresponding SB limits. This will allow one to compare with lattice data above $T_c$ \cite{49,52} as well.

\begin{figure}
\begin{center}
\includegraphics[width=10cm]{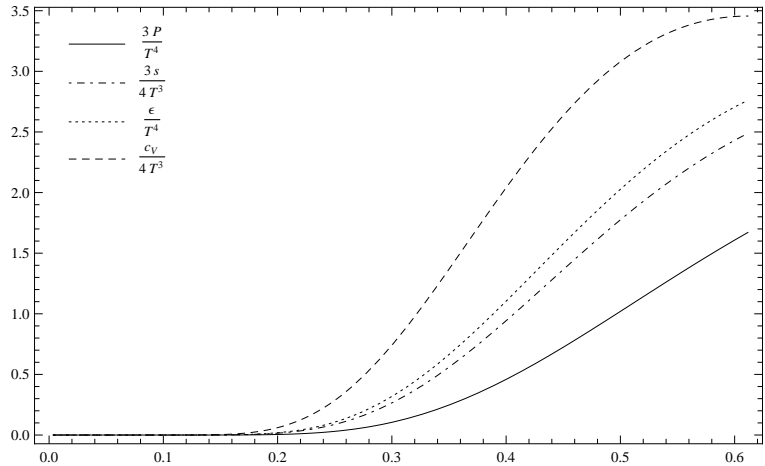}
\caption{The NP gluon pressure, the entropy and energy densities, the heat
capacity as functions of $T/T_c$ in the low-temperatures region
shown up to $0.6T_c$.}
\label{fig:6}
\end{center}
\end{figure}

\begin{figure}
\begin{center}
\includegraphics[width=10cm]{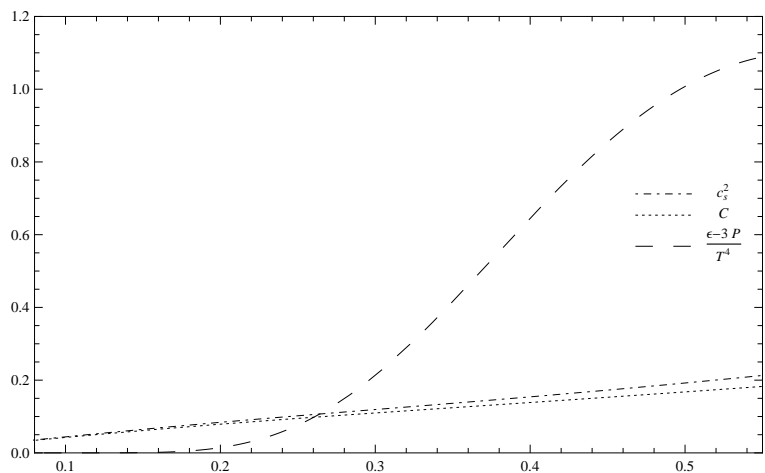}
\caption{The NP velocity of sound, conformity and the trace
anomaly relation as functions of $T/T_c$. They are shown
in the low-temperatures region up to $0.6T_c$.}
\label{fig:7}
\end{center}
\end{figure}

\begin{figure}
\begin{center}
\includegraphics[width=10cm]{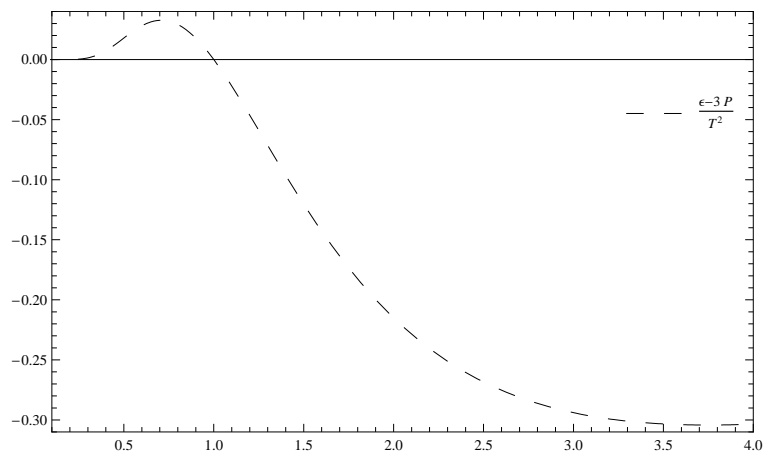}
\caption{The NP trace anomaly relation scaled by $T^2$ in $\GeV^2$ units (see Eq.~(48)) as a function of $T/T_c$.
It clearly approaches a finite constant in the high temperature limit (compare with Fig. 4).}
\label{fig:8}
\end{center}
\end{figure}

\begin{figure}
\begin{center}
\includegraphics[width=10cm]{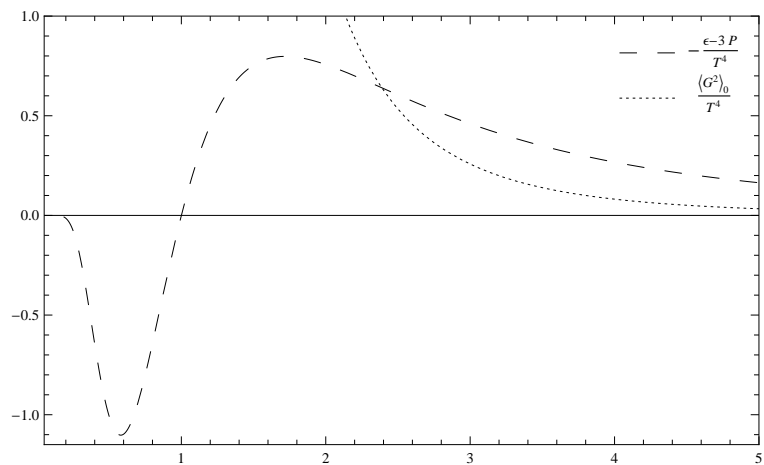}
\caption{The zero temperature gluon condensate and the NP trace anomaly relation with minus sign, as it enters Eq.~(49). Both scaled by $T^4$ and shown as functions of $T/T_c$.}
\label{fig:9}
\end{center}
\end{figure}

\begin{figure}
\begin{center}
\includegraphics[width=10cm]{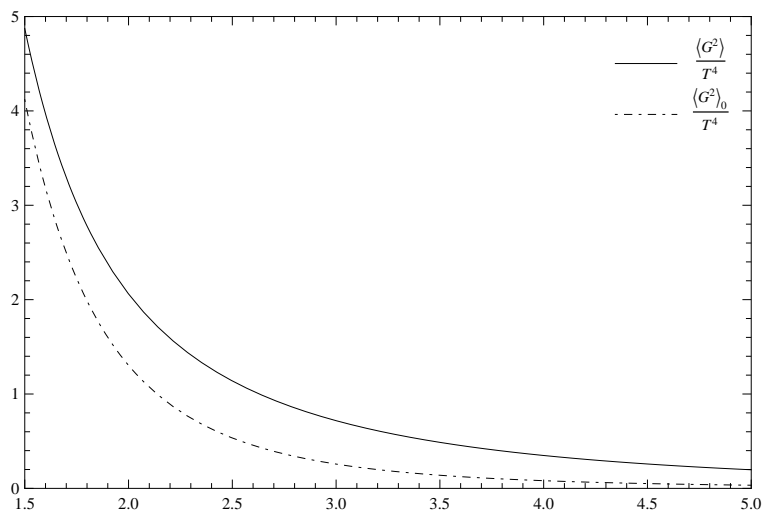}
\caption{The NP gluon condensate scaled by $T^4$ as a function of $T/T_c$.
The zero temperature gluon condensate scaled by $T^4$ is also shown.
At temperatures above $1.5T_c$ they go down as explicitly shown by Eq.~(49). }
\label{fig:10}
\end{center}
\end{figure}

\subsection{The mass gap and a "fuzzy bag" term }

On account of the first of Eqs.~(28) and Eq.~(31), it is instructive to re-write the NP gluon
pressure (30) equivalently as follows:

\begin{equation}
P_{NP}(T) = \Delta^2 T^2 - {6 \over \pi^2} \Delta^2 P'_1 (T) + {16 \over \pi^2} T
[P_2(T) + P_3(T) - P_4(T)],
\end{equation}
where

\begin{equation}
P'_1(T) = \int^{\omega_{eff}}_0 \dd \omega {\omega \over e^{\beta\omega} -1}.
\end{equation}
It scales as $T \omega_{eff}$ in the $T \rightarrow \infty$ limit, while the integrals $P_2(T)$, $P_3(T)$ and $P_4(T)$ remain unchanged, see Eqs.~(32). Let us emphasize the explicit presence of the NP mass gap term in Eq.~(46), namely
$\Delta^2 T^2$. It will be present in the full EoS as well, since in the PT part of the gluon pressure
the mass gap contribution is logarithmical suppressed (see Eq.~(33)), and hence its contribution cannot be canceled. This is in full agreement with the so-called "fuzzy bag" model \cite{52,54,55,56} and
with the massless Boltzmann stringy model \cite{57}. The presence of the mass gap $\Delta^2$ in our approach from the very beginning naturally explains the properties of such models.

The NP trace anomaly relation scaled by $T^2$, t.e.,

\begin{equation}
{\epsilon_{NP}(T)  - 3 P_{NP}(T) \over T^4} \times T^2 = {\epsilon_{NP}(T)  - 3 P_{NP}(T) \over T^2} \longrightarrow - const., \ T \rightarrow \infty
\end{equation}
is shown in Fig. 8. It clearly goes to a finite constant (having the dimensions of $\Delta^2$)
in the high temperature limit, as it should
in accordance with the previous discussion and the thermal lattice QCD results \cite{49,52,55,58}.

This effect immediately shows up in the NP gluon condensate (39). Scaled by $T^4$, analytically it becomes

\begin{equation}
{<G^2>_T \over T^4} = { <G^2>_0 \over T^4}  - {[ \epsilon_{NP}(T) - 3P_{NP}(T)] \over T^2} \times { 1 \over T^2}
\sim  { const. \over T^2}, \ T \rightarrow \infty.
\end{equation}
The difference in the fall off at high temperatures between the zero temperature gluon condensate
and minus trace anomaly relation (as it enters Eq.~(49)), both scaled by $T^4$, is clearly seen in Fig. 9.
Evidently, this difference is due to the asymptotic behavior of Eq.~(48). At temperatures below $T_c$ there is no difference between $<G^2>_T / T^4$ and $<G^2>_0 / T^4$, which is as it should be (see, for example again Fig. 5).
At temperatures above $T_c$ the difference between these two quantities is explicitly determined by Eq.~(49) and
is clearly shown in Fig. 10. The presence of the mass gap contribution
in Eq.~(46), and hence in the trace anomaly relation (38) scaled by $T^4$, explains that the way it goes down
is not $1/T^4$ \cite{49,58} but, as follows from our consideration, it decreases as $1/T^2$.

The inclusion of the PT contribution $[\epsilon_{PT}(T)  - 3 P_{PT}(T)] / T^4$ in Eq.~(49) will produce a rather small numerical correction at high temperatures due to the SB limit (42) for the trace anomaly. The numerical results presented in Table 1 confirm this statement. It shows rather good numerical agreement between our NP analytical trace anomaly, as it enters Eq.~(49), and its lattice counterpart at hight temperatures starting already from $2T_c$ \cite{49}.

In conclusion, let us note that in lattice QCD at non-zero temperatures there is a possibility to
distinguish between the magnetic (m) and electric (e) parts of the gluon condensate
(see, for example \cite{49,59,60} and references therein). In \cite{59,60} and \cite{49}, such a gauge-invariant field-strength correlators in pure YM theory have been calculated at finite temperatures
around $T_c$ and above $T_c$, respectively. Following \cite{60}, we can formally write
$<G^2>_T = <G^2_e>_T + <G^2_m>_T$, but at this stage we do not know how to calculate these two parts separately within our approach. There is therefore as yet no possibility to compare with lattice data. In any case, such an analytical investigation at zero and non-zero temperatures is beyond the scope of the present investigation.

\begin{table}[h!b!p!]
\caption{The NP analytical and lattice trace anomalies at high temperatures.}
\begin{center}
\begin{tabular}{ c | c | c | }
\cline{2-3}
& \multicolumn{2}{|c|}{Numbers} \\ \cline{1-3}
\multicolumn{1}{|c|}{${T}/{T_{c}}$} & Ours, \ Fig. 9 & Lattice, reference \cite{49} \\ \hline \hline
\multicolumn{1}{|c|}{$2$} & $0.762$ & $0.860$ \\ \hline
\multicolumn{1}{|c|}{$3$} & $0.453$ & $0.475$ \\ \hline
\multicolumn{1}{|c|}{$4$} & $0.252$ & $0.255$ \\ \hline
\multicolumn{1}{|c|}{$5$} & $0.166$ & $0.166$ \\ \hline
\end{tabular}
\end{center}
\end{table}

\subsection{The NP dynamical structure of the GM. A brief description}

The NP dynamical structure of the GM within our approach is highly non-trivial. In the whole temperature
range we have the two different massive gluonic excitations
$\omega'$ and $\bar \omega$ with the effective masses $m'_{eff}=
1.17 \ \GeV$ and $\bar m_{eff} = 0. 585 \ \GeV$, respectively. Both effective
masses are due to the mass gap $\Delta^2$, which is responsible
for the large-scale structure of the QCD ground state. It is dynamically generated by the
nonlinear interaction of massless gluon modes \cite{43,45,46}. The first
massive excitations can be interpreted as the glueballs, since $m'_{eff}$ is comparable to the masses
of scalar glueballs (though it is lighter than the pseudoscalar ones) \cite{61}. The second one
$\bar m_{eff}$ might be identified
with an effective gluon mass of about $(500-800) \ \MeV$, which arises in different approaches (see again the
above-mentioned review \cite{61} and references therein). We also have the two different massless gluonic
excitations $\omega$: the NP massless gluons, propagating
according to the integral $P_1(T)$ in Eq. (31), and almost SB
massless gluons, propagating according to the integral $P_2(T)$
in Eqs. (32). We stress that the integral (31) should be multiplied by $(6 /\pi^2) \Delta^2$, and all other integrals (32) are to be multiplied by $(16 / \pi^2) T$, when one speaks about different NP contributions to the NP pressure (30).

It is worth emphasizing once more that all these NP gluonic excitations are of dynamical origin, and only such
can be accounted for within our approach. Whether the PT part (33) of the gluon pressure (29) will introduce a new massless or even massive gluonic excitations (or, equivalently, an effective gluonic degrees of freedom \cite{16,62}) is an open question at this stage. In our opinion, however, they may be really present in
the full GM EoS. In addition to the mass gap the PT part (33) depends on the asymptotic scale parameter of QCD.
The combination of these two fundamental scale parameters may lead to the creation of
new massive gluonic excitations around $T_c$, for example, pseudoscalar glueballs \cite{60,61,63}.
In any case, we should definitely have the SB free massless gluons far away from $T_c$.
However, one thing should be made perfectly clear. Because of these and a possible new massive excitations
in the GM, the behavior of all the thermodynamic variables in the moderate temperatures region $(1-3.75)T_c$ will be substantially different from the behavior of a gas of free massless gluons. This our qualitative yet conclusion is in
agreement with recent lattice data and quasi-particle approaches both cited above. The recent lattice data
will be reproduced quantitatively after the evaluation of the NP (30), PT (33), and free gluon contributions to the full GM pressure.

The existence of the SB-type relations (43)-(44) indicates that in
the transition region near $T_c$ a dramatic increase in the number of effective gluonic
degrees of freedom will appear (for example, the glueballs will begin rapidly to dissolve).
This effect will be reflected in the strong increase of the thermodynamic variables near $T_c$, in agreement with
recent lattice results \cite{60,63,64}.
This will lead to drastic changes in the structure of the GM. A change in this number is enough
to generate pressure gradients, but not enough to affect the
pressure itself. It varies slowly and therefore remains continuous
in this region. At the same time, the pressure gradients, such as
the energy and entropy densities, etc., should undergo sharp changes in their behavior.
The penetration of the PT part (33) of the gluon pressure below $T_c$ should
not go deeply into the transition region $(0.6-1)T_c$, since the INP effective
charge is logarithmical suppressed. However, it should go
sufficiently deep in order to eliminate "non-physical" maxima
and other effects, created by the derivatives of $P_{NP}(T)$ in the transition region.
Instead of "non-physical" maxima, the full thermodynamic
quantities (apart from the pressure) should have
discontinuities at $T_c$, as follows from our approach. This has also
been established by the thermal QCD lattice cited above.

Of course, not all the glueballs will be dissolved in the transition
region. Some of them will remain above $T_c$, along with other
massive and massless gluonic excitations, forming thus a possible
mixed and extended mixed phases around $T_c$ \cite{53}.
After the evaluation of the PT part (33), which is also "contaminated" by the NP contributions, we hope
that the NP physics of the mixed phases will be well understood in the GM.
At very high temperatures, starting at $3.75T_c$ the NP effects become very small,
and the structure of the GM will be mainly determined by the SB relations (40)-(42) between
all the important thermodynamic quantities.

\section{Conclusions}

The effective potential approach for composite operators \cite{26}
has been generalized to non-zero temperatures in order to derive
analytical EoS for pure $SU(3)$ YM fields, shown in Eqs.~(29)-(33).
In its NP part (30) there is no dependence on the coupling constant, only
the dependence on the mass gap, which is responsible for the
large-scale structure of the QCD ground state. A key element of
this work is the generalization of the expression for the bag constant at zero
temperature \cite{27} to non-zero temperatures. The NP part (30) of the gluon pressure (29) has been exactly evaluated, while its PT part (33) has been left undetermined at this stage, since it requires a separate investigation.

Our main quantitative and qualitative results in this investigation, which will not be changed
(or only slightly changed) after the inclusion of the PT part (33), are:

{\bf (i)}. In Eq.~(29) the confining dynamics at non-zero temperatures (15) is taken into account through
the $T$-dependent bag constant (22) and the mass gap $\Delta^2$.

{\bf (ii)}. Other NP effects are also taken into account via the YM part (27).

{\bf (iii)}. The mass gap $\Delta^2$ or, equivalently, $\omega_{eff}$ is the only
one independent input scale parameter needed to calculate the NP thermodynamic quantities.

{\bf (iv)}. The presence of the four different types of massive and massless
gluonic excitations of the NP origin.

{\bf (v)}. The characteristic temperature $T_c = 266.5 \ \MeV$
is a temperature at which the maximum of the NP part of the gluon pressure is achieved.

{\bf (vi)}. The low-temperatures region up to $0.6 T_c$ is under control. The exponential fall off in the
$T \rightarrow 0$ limit of all the main NP thermodynamic quantities (see Fig. 6) is expected to be preserved
for their full counterparts as well.

{\bf (vii)}. The behavior of all the thermodynamic quantities in the transition region $(0.6-1)T_c$
           depends on how deeply the PT part (33) penetrates this region.

{\bf (viii)}. The existence of the SB-type relations (43)-(44) at $T_c$  shows that the structure of
the GM below and above $T_c$  may be really rather different.

{\bf (ix)}. Since the NP entropy and energy densities satisfy them, we expect for their full counterparts
to have jump discontinuities at $T_c$, while the full pressure remains continuous.

{\bf (x)}. Since the NP heat capacity does not satisfy them, we expect for its full counterpart to have
an essential discontinuity at $T_c$, while the full speed of sound squared is to be zero at $T_c$.

{\bf (xi)}. In the moderate temperatures region $(1-3.75)T_c$ the NP vacuum effects are still significant.

{\bf (xii)}. Because of this, our qualitative prediction is that in this region the behavior of all
the thermodynamic variables will be substantially different from the behavior of a gas of free massless gluons.

{\bf (xiii)}. All the full thermodynamic quantities, therefore, approach their SB limits rather slowly.

{\bf (xiv)}. A possible understanding of new forms of the GM below and above $3.75T_c$ as strong and
weak coupling GM, respectively, has to be pointed out.

{\bf (xv)}. The existence of the mass gap term $\Delta^2 T^2$ in the NP pressure (46),
which remains in the gluon pressure (29) as well.

{\bf (xvi)}. Because of the mass gap term $\Delta^2 T^2$ the full trace anomaly and the
gluon condensate will go down as $1/T^2$ at high temperatures, and not as $1/T^4$.

\vspace{3mm}

In the subsequent paper (part II) we shall evaluate the PT part (33) of the gluon pressure (29), as well as include the free gluons contribution. It will make it possible to establish the order of the phase transition at $T_c$. We will be able to compare our numerical results with thermal QCD lattice calculations \cite{49,52}
at high temperatures above $T_c$ and in the transition region \cite{52,60,63} as well.
Only after completion of this programme, we will include the quark degrees of freedom
in order to derive the NP QGP EoS within our formalism. For example, this will allow one to confirm  a possible existence of Quarkyonic Matter (QM) \cite{65,66} and of a triple point in the QCD phase diagram
(see \cite{67} and references therein). Finally, let us note that recently \cite{68} the pressure and the trace anomaly relation both scaled by $T^4$ have been calculated in 3d "electrostatic QCD" in apparent contradiction with the 4d lattice data \cite{49,52}. It is worth noting once more that our calculations are in a good agreement with them (see Table 1 and discussion above).

\ack{
This paper is dedicated to the memory of the late Prof. J. Zimanyi, who has initiated this investigation.
Support by HAS-JINR Scientific Agreement (P. Levai) is to be acknowledged. We would
like to thank L. Csernai, T. Bir\'{o}, T. Csorg\"o, Gy. Wolf, P. Van, G. Barnaf\"oldi, V. Skokov and especially C. Wilkin and S. Pochybova for useful discussions, comments, remarks and help. One of the authors (V.G.) is grateful
to V.K. and A.V. Kouzushins for constant support and interest.}

\appendix
\section{The summation of the thermal logarithms}

In the second terms of Eqs. (19) and (25) the summation over
the Matsubara frequencies can be done explicitly \cite{7}, as follows:

\begin{eqnarray}
\sum_{n= - \infty}^{+ \infty} { 1 \over \omega^2 + \omega^2_n} &=&
\sum_{n= - \infty}^{\infty} { 1 \over \omega^2 + (2\pi T)^2 n^2} =
\left({\beta \over 2 \pi} \right)^2 \sum_{n= -
\infty}^{ + \infty} { 1 \over n^2 + (\beta \omega / 2 \pi)^2} \nonumber\\
&=& \left( {\beta \over 2 \pi} \right)^2 {2 \pi^2 \over \beta \omega} \left( 1 + { 2
\over e^{\beta\omega} -1} \right)=  { \beta \over 2 \omega} \left(
1 + { 2 \over e^{\beta\omega} -1} \right).
\end{eqnarray}

In terms of the above-introduced parameters, the sums in Eq. (19)
containing the corresponding logarithms look like:

\begin{equation}
\sum_{n= - \infty}^{+ \infty} \ln [ 3 \Delta^2 + \omega^2 +
\omega^2_n] = \ln \omega'^2 + 2 \sum_{n= 1}^{ \infty} \ln (2\pi /
\beta)^2[ n^2 + (\beta \omega' / 2\pi)^2]
\end{equation}
and

\begin{equation}
\sum_{n= - \infty}^{+ \infty} \ln [\omega^2 + \omega^2_n] = \ln
\omega^2 + 2 \sum_{n= 1}^{ \infty} \ln (2\pi / \beta)^2[ n^2 +
(\beta \omega / 2\pi)^2].
\end{equation}

It is convenient to introduce the notations:

\begin{equation}
L(\omega') = \sum_{n=1}^{ \infty} \ln[ n^2 + (\beta \omega' /
2\pi)^2] = \sum_{n=1}^{ \infty} \ln n^2  + \sum_{n= 1}^{ \infty}
\ln \left[ 1 - {x'^2 \over n^2 \pi^2} \right]
\end{equation}
and similarly, letting $\omega' \rightarrow \bar \omega$

\begin{equation}
L(\omega) = \sum_{n=1}^{ \infty} \ln[ n^2 + (\beta \omega /
2\pi)^2] = \sum_{n=1}^{ \infty} \ln n^2  + \sum_{n= 1}^{ \infty}
\ln \left[ 1 - { x^2 \over n^2 \pi^2} \right].
\end{equation}
In these expressions we introduced the following notations:

\begin{equation}
x'^2  = - \left( {\beta \omega' \over 2 } \right)^2, \quad x^2 = -
\left( {\beta \omega \over 2 } \right)^2.
\end{equation}
So the difference $L(\omega') - L(\omega)$ becomes

\begin{eqnarray}
L(\omega') - L(\omega) &=& \sum_{n= 1}^{ \infty} \ln \left[ 1 -
{x'^2 \over n^2 \pi^2} \right] - \sum_{n= 1}^{ \infty} \ln \left[
1 - { x^2 \over n^2 \pi^2} \right] \nonumber\\
&=& \ln \sin x' - {1 \over 2} \ln
x'^2 - \ln \sin x + {1 \over 2} \ln x^2,
\end{eqnarray}
or, equivalently,

\begin{equation}
L(\omega') - L(\omega) = - {1 \over 2} \ln \left( { x'^2 \over
x^2} \right) + \ln \left( {\sin x'  \over \sin x} \right).
\end{equation}
From the relation (A.6) it follows that

\begin{equation}
x'  = \pm i \left( {\beta \omega' \over 2 } \right), \quad x = \pm
i \left( {\beta \omega \over 2 } \right),
\end{equation}
so eq.~(A.8) finally becomes

\begin{equation}
 L(\omega') - L(\omega) = - { 1 \over 2} \ln \left( {
\omega'^2 \over \omega^2} \right) + { 1 \over 2} \beta( \omega' -
\omega) + \ln \left({1 - e^{- \beta \omega'} \over 1 - e^{- \beta
\omega}} \right).
\end{equation}

\section{The scale-setting scheme}

Let us note that $\omega_{eff}$, which appears first
in the integral (21), is the only free parameter of our
approach. In frequency-momentum space it is

\begin{equation}
\omega_{eff} = \sqrt{q^2_{eff}  - \omega^2_c},
\end{equation}
where we introduce the "constant" Matsubara frequency $\omega_c$,
which is always positive. Hence $\omega_{eff}$ is always less or
equal to the $q_{eff}$ of four-dimensional QCD, i.e., $\omega_{eff}
\leq q_{eff}$. One can then conclude that $q_{eff}$ is a very good
upper limit for possible values of $\omega_{eff}$. In this
connection, let us recall that the bag constant $B_{YM}$ at zero
temperature has been successfully calculated at a scale $q^2_{eff}
= 1 \ \GeV^2$, in fair agreement with other phenomenological
quantities such as gluon condensate \cite{27}. So $\omega_{eff}$
is fixed as follows:

\begin{equation}
\omega_{eff} = \sqrt{q^2_{eff}} = 1 \ \GeV.
\end{equation}
The mass gap squared $\Delta^2$, also calculated at this scale, has a value \cite{27}

\begin{equation}
\Delta^2 = 0.4564 \ \GeV^2, \quad \Delta = 0.6756 \ \GeV.
\end{equation}
The effective gluon masses, defined in the relations (20) and
(26), then become

\begin{equation}
m'_{eff} = \sqrt{3} \Delta = 1.17 \ \GeV, \quad \bar m_{eff} =
{\sqrt{3} \over 2} \Delta = 0.585 \ \GeV.
\end{equation}

The above-mentioned gluon condensate at zero temperature $<G^2>_0$ calculated at the scale (B.2)
and at the same confining effective charge (13) in \cite{27} is

\begin{equation}
<G^2>_0 \equiv  \langle{0} | {1 \over 4}G^a_{\mu\nu} G^a_{\mu\nu} | {0}\rangle  = 0.1052 \ \GeV^4.
\end{equation}
We need this value for the calculation of the temperature-dependent gluon condensate $<G^2>_T$ via Eq.~(39).
Multiplying the right-hand-side of the relation (B.5) by $( 4 \alpha_s / \pi)$, where
$\alpha_s=0.1187$ \cite{69}, it numerically becomes

\begin{equation}
\langle{0} | { \alpha_s \over \pi }G^a_{\mu\nu} G^a_{\mu\nu} | {0}\rangle \approx 0.016 \ \GeV^4.
\end{equation}
It is in a good agreement with its phenomenological value,
$<G^2>_{ph} \equiv  \langle{0} | (\alpha_s / \pi )G^a_{\mu\nu} G^a_{\mu\nu} | {0}\rangle \approx 0.012 \ \GeV^4$,
which can be changed by a factor of $\sim 2$, as mentioned in \cite{41} (see also \cite{70} and references therein). In this connection, we recall  \cite{27} that the quark contribution to the bag constant, and hence to the gluon condensate, is approximately
an order of magnitude less than the pure YM one. So the above-mentioned agreement of our YM value (B.6) with the phenomenological value is impressive.

\end{document}